\documentclass[twocolumn,aps,prl] {revtex4} 
\usepackage{graphicx}
\begin{document}
\pagestyle{empty} 
\title{Fluid flow at the interface between elastic solids with randomly rough surfaces}
\author{  
B.N.J. Persson} 
\affiliation{IFF, FZ J\"ulich, D-52425 J\"ulich, Germany}

\begin{abstract}
I study fluid flow at the interface between elastic solids with randomly rough surfaces.
I use the contact mechanics model of Persson to take into account the elastic interaction
between the solid walls and the Bruggeman effective medium theory to account for the influence of the disorder on the fluid flow.
I calculate the flow tensor which determines the pressure flow factor and, e.g., the leak-rate of static seals.
I show how the perturbation treatment of Tripp
can be extended to arbitrary order in  the ratio between the root-mean-square 
roughness amplitude and the average interfacial surface
separation.  I introduce a matrix $D(\zeta)$, determined by the surface roughness power spectrum,
which can be used to describe the anisotropy of the surface at any magnification $\zeta$. 
I present results for the asymmetry factor $\gamma (\zeta)$ (generalized Peklenik number) for 
grinded steel and sandblasted PMMA surfaces. 

\end{abstract}
\maketitle


\vskip 0.3cm
{\bf 1. Introduction}

The influence of surface roughness on fluid flow at the interface between solids in stationary or sliding contact is
a topic of great importance both in Nature and Technology. Technological applications includes leakage of seals,
mixed lubrication, and removal of water from the tire-road footprint. In Nature fluid removal (squeeze-out) 
is important for adhesion and grip between the tree frog or Gecko adhesive toe pads and the countersurface during raining,
and for cell adhesion. 

Almost all surfaces in Nature and most surfaces of interest in Tribology have roughness on many different length
scales, sometimes extending from atomic distances ($\sim 1 \ {\rm nm}$) to the macroscopic size of the system which could be of order
$\sim 1 \ {\rm cm}$. Often the roughness is fractal-like so that when a small region is magnified (in general with different
magnification in the parallel and orthogonal directions) it ``looks the same'' as the unmagnified surface. 

Most objects produced in engineering have some particular macroscopic shape characterized by a radius of curvature 
(which may vary over the surface of the solid) e.g., the radius $R$ of a cylinder in an engine. 
In this case the surface may appear perfectly smooth
to the naked eye but at short enough length scale, in general much smaller than $R$, the surface will exhibit strong irregularities
(surface roughness). The surface roughness power spectrum $C({\bf q})$ of such as surface will exhibit a roll-off wavelength
$\lambda_0 << R$ (related to the roll-off wavevector $q_0=2 \pi /\lambda_0$) and will appear smooth (except for the macroscopic curvature $R$)
on length scales much longer than $\lambda_0$. In this case, when studying the fluid flow between two macroscopic solids, one may
replace the microscopic equations of fluid dynamics with effective equations 
describing the average fluid flow on length scales much larger than $\lambda_0$, and which can be used to study, e.g., the lubrication of the
cylinder in an engine. This approach of eliminating or integrating out short length scale degrees of freedom to obtain effective
equations of motion which describes the long distance (or slow) behavior is a very general and powerful concept often used in Physics.
 
In the context of fluid flow at the interface between closely spaced solids with surface roughness, Patir and Cheng\cite{PC1,PC2} 
have showed how the Navier-Stokes equations of fluid dynamics
can be reduced to effective equations of motion involving locally averaged fluid pressure and flow velocities.
In the effective equation occur so called flow factors, which are functions of the locally averaged interfacial
surface separation $\bar u$. They showed how the flow factors can be determined by solving numerically the fluid flow in small rectangular
units with linear size of order (or larger than) 
the roll-off wavelength $\lambda_0$ introduced above. However, with the present speed (and memory) limitations of computers 
fully converged solutions using this approach can only take into account roughness over two or at most three decades in length scale. 
In addition, Patir and Cheng did not include the long-range elastic deformations of the solid walls in the analysis. Later studies have attempted to include
elastic deformation using the contact mechanics model of Greenwood-Williamson (GW)\cite{GW}, but it is now known that this theory 
(and other asperity contact models \cite{Bush}) 
does not correctly describe contact mechanics because of the neglect of the long range elastic coupling
between the asperity contact regions\cite{PerssonJPCM,Carlos}. In particular, the relation between the average interfacial separation $\bar u$ and the
squeezing pressure $p$, which is very important for the fluid flow problem, is incorrectly described by the GW model
[the GW model predict asymptotically (for large $\bar u$) $p\sim {\rm exp}(-a \bar u^2)$, while the exact result\cite{PerssonPRL,YangPersson,Lorenz} for randomly rough surfaces
is $p\sim {\rm exp}(-b \bar u)$, where $a$ and $b$ are constants determined by the nature of the surface roughness].

The paper by Patir and Cheng was followed by many other studies of how to eliminating or integrate out the surface roughness 
in fluid flow problems (see, e.g., the work by Sahlin et al.\cite{Salin}). Most of these theories involves solving numerically for the fluid flow
in rectangular interfacial units and, just as in the Patir and Cheng approach,
cannot include roughness on more than $\sim 2$ decades in length scale. In addition, in some
of the studies the measured roughness topography must be ``processed'' 
in a non-trivial way in order to obey periodic boundary conditions
(which is necessary for the Fast Fourier Transform  method used in some of these studies). 

Tripp\cite{Tripp} has presented an analytical derivation of the flow factors for the case where the separation between the surfaces is so large
that no direct solid-solid contact occurs. He obtained the flow factors to first order in $\langle h^2\rangle /\bar u^2$, where
$\langle h^2\rangle$ is the ensemble average of the square of the roughness amplitude and $\bar u$ the average surface separation. This result is of
great conceptual importance, but of minor practical importance as the influence of the surface roughness on the fluid flow becomes important only
when direct solid-solid contact occur. 

Many surfaces of practical importance have roughness with isotropic statistical properties, e.g., sandblasted surfaces or surfaces coated with particles typically bound
by a resin to an otherwise flat surface, e.g., sandpaper surfaces. However some surfaces of engineering interest have
surface roughness with anisotropic statistical properties, e.g., surfaces which have been polished or grinded in one direction. The
theories of Patir and Chen\cite{PC1,PC2} and of Tripp\cite{Tripp} can be applied also to surfaces with
anisotropic statistical properties.
The surface anisotropy is usually characterized by a single number, the so called Peklenik number $\gamma$, which is the ratio between the decay length
of the height-height correlation function $\langle h(x,y)h(0,0) \rangle$ along the $x$ and $y$-directions, i.e.,
$\gamma = \xi_x/\xi_y$ where $\langle h(\xi_x,0)h(0,0) \rangle = \langle h(0,0)h(0,0) \rangle /2$ 
and $\langle h(0,\xi_y)h(0,0) \rangle = \langle h(0,0)h(0,0) \rangle /2$. Here it has been assumed that the $x$-axis is oriented along one
of the principal direction of the anisotropic surface roughness. However, the anisotropy properties of a surface may depend on the resolution
(or magnification) which is not taken into account in this picture.  

In this paper I will present a new approach to calculate the fluid flow at the interface between two elastic solids
with randomly rough surfaces. The present treatment is based on a recently developed theory for calculating the leak rate of
stationary seals\cite{LP}. The theory use the contact mechanics theory of Persson\cite{JCPpers,PSSR} in combination
with the Bruggeman effective medium theory to calculate the fluid conductivity tensor. In this paper we will generalize the
treatment presented in Ref. \cite{LP} to surfaces with random roughness with anisotropic statistical properties. We also introduce
a generalized Peklenik number $\gamma (\zeta)$ which depends on the magnification $\gamma$. Thus the theory takes
into account that the anisotropy properties of the surface roughness may depend on the magnification 
under which the surface is observed. We present results for how $\gamma (\zeta)$ depends on $\zeta$ for a grinded steel
surface studied using Atomic Force Microscopy and Scanning Tunneling Microscopy, and for a sandblasted PMMA surface studied
using an optical technique. As an illustration we calculate the pressure flow factor for surfaces with anisotropic properties.
We emphasize that the present treatment accurately accounts for surface roughness on arbitrary many decades in
length scale, and a full calculation typically takes less than a minute on a normal PC. 
In particular, the presented theory should be very useful
for gaining a quick insight into what are the most important length scales in the problem under study. 

This paper is organized as follows: In Sec. 2 we briefly review the basic equations of fluid dynamics and describe
some simplifications which are valid in the present case. In Sec. 3 and Appendix A I show how the perturbation treatment of Tripp
can be extended to arbitrary order in $\langle h^2 \rangle /\bar u^2$. This treatment may not be so important for the
fluid flow problem we consider as it is necessary to take into account that asperity contact occur
already for relative small values of $\langle h^2 \rangle /\bar u^2$, but the approach may find applications in other contexts.
In addition, the solution we present in wavevector space differ from the treatment of Tripp and leads directly to
a matrix $D(\zeta)$ which we used to describe the anisotropy of the surface at any magnification $\zeta$. In Sec.
4 we define $D(\zeta)$ and present results for how the asymmetry factor $\gamma (\zeta)$ (generalized Peklenik number) depends
on the magnification $\zeta$. 
In Sec. 5 we briefly review the contact mechanics model we use. In Sec. 6 we describe the critical junction
theory for the flow factor, and in Sec. 7 and 8 we show how the Bruggeman effective medium theory can
be used in combination with the contact mechanics theory to calculate the fluid flow tensor which determines the pressure flow
factor and, e.g., the leak-rate of stationary seals. Sec. 9 contains the summary.

\begin{figure}
\includegraphics[width=0.45\textwidth,angle=0.0]{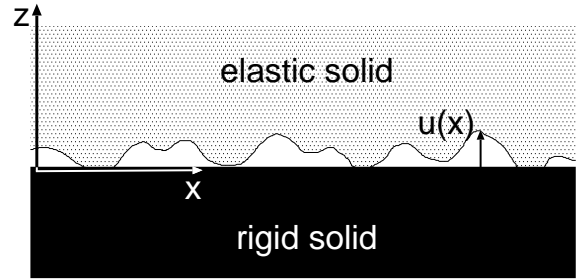}
\caption{\label{geometry}
An elastic solid with a rough surface in contact with a rigid solid with a flat surface.
}
\end{figure}

\vskip 0.3cm
{\bf 2. Fluid flow between solids with random surface roughness}

Consider two elastic solids with randomly rough surfaces. Even if the solids are squeezed in contact,
because of the surface roughness there will in general be non-contact regions at the interface and, if the
squeezing force is not too large, there will exist non-contact channels from one side to the other side of the nominal
contact region. We consider now fluid flow at the interface between the solids. We assume that the fluid is Newtonian and
that the fluid velocity field ${\bf v}({\bf x},t)$ satisfies the Navier-Stokes equation:
$${\partial {\bf v} \over \partial t} + {\bf v}\cdot \nabla {\bf v} = -{1\over \rho} \nabla p + \nu \nabla^2 {\bf v}$$  
where $\nu = \eta /\rho$ is the kinetic viscosity and $\rho$ the mass density. For simplicity we will also assume
an incompressible fluid so that
$$\nabla \cdot {\bf v} = 0$$
We assume that the non-linear term ${\bf v}\cdot \nabla {\bf v}$ can be neglected (which correspond to
small inertia and small Reynolds number), which is usually the case in fluid flow between narrowly spaced solid walls. 
For simplicity we assume the lower solid to be rigid with
a flat surface, while the upper solid is elastic with a rough surface. 
Introduce a coordinate system $xyz$ with the
$xy$-plane in the surface of the lower solid and the $z$-axis pointing towards the upper solid, see Fig. \ref{geometry}. 
The upper solid moves with
the velocity ${\bf v}_0$ parallel to the lower solid. 
Let $u(x,y,t)$ be the separation between the solid walls and assume that the slope $|\nabla u| << 1$. 
We also assume that $u/L << 1$, where $L$ is the linear size of the nominal contact region. In this case one expect that the fluid velocity varies slowly
with the coordinates $x$ and $y$ as compared to the variation in the orthogonal direction $z$.
Assuming a slow time dependence the Navier Stokes equations reduces to
$$\eta {\partial^2 {\bf v} \over \partial z^2} = \nabla p$$
Here and in what follows ${\bf v} = (v_x,v_y)$, ${\bf x}=(x,y)$ and $\nabla = (\partial_x,\partial_y)$ are two-dimensional 
vectors. Note that $v_z \approx 0$ and that $p({\bf x})$ is independent of $z$ to a good approximation.
The solution to the equations above can be written as
$${\bf v} = {1\over 2 \eta} z(z-u({\bf x}))\nabla p + {z\over u({\bf x})}{\bf v}_0$$
so that ${\bf v}=0$ on the solid wall $z=0$ and ${\bf v}={\bf v}_0$ for $z=u({\bf x})$.
Integrating over $z$ (from $z=0$ to $z=u({\bf x})$)
gives the fluid flow vector
$${\bf J} = - {u^3({\bf x})\over 12 \eta}\nabla p +  {1\over 2} u({\bf x}) {\bf v}_0\eqno(1)$$
Mass conservation demand that
$${\partial u({\bf x},t) \over \partial t} + \nabla \cdot {\bf J} = 0$$ 
where the interfacial separation $u({\bf x},t)$ is the volume of fluid per unit area. In this last equation we have allowed
for a slow time dependence of $u({\bf x},t)$ as would be the case, e.g., during fluid squeeze-out from the 
interfacial region between two solids. However, in this paper we will only focus on the case where 
$u$ is time independent so that $\nabla \cdot {\bf J} = 0$. This case is relevant for, e.g., fluid
leakage in stationary seals.

\vskip 0.3cm

{\bf 3. Perturbation treatment}

Here we show how one can obtain an effective flow equation by integrating out the short-wavelength roughness.
We first re-derive the (first order) expansion result of Tripp in wavevector space. After that we present the results of a Renormalization
Group type of treatment (the derivation is presented in Appendix A).
The treatment presented here does not take into account the elastic interaction between the solid walls and is therefore
strictly valid only for large enough average wall-wall separation.

Let $u({\bf x}) = \bar u +h({\bf x})$ denote the local surface separation, where $\bar u = \langle u \rangle$ is the average separation
($\langle .. \rangle$ stands for ensemble averaging), 
and $h({\bf x})$ is the contribution from the surface roughness with $\langle h \rangle = 0$. In this section we assume
$h/\bar u << 1$ and perform a perturbation expansion in the small parameter $h/\bar u$. Let us write the fluid pressure as
$$p=p_0+p_1+p_2+...$$
where $p_0$ is the pressure to zero order in $h$ (so that $p_0 = \langle p_0 \rangle$), 
$p_1$ to first order in $h$ and so on. The fluid flow current is given by
$${\bf J} = -{u^3\over 12 \eta} \nabla p +{1\over 2}u {\bf v}$$
Thus to second order in $h$ we get
$${\bf J} = - {\bar u^3\over 12 \eta} \nabla (p_0+p_1+p_2)$$
$$- {3\bar u^2 h \over 12 \eta}\nabla (p_0 +p_1) 
- {3\bar u h^2 \over 12 \eta}\nabla p_0 +{1\over 2}(\bar u +h){\bf v}\eqno(2)$$
The ensemble average of this equation gives
$$\langle {\bf J} \rangle = - {\bar u^3\over 12 \eta} \nabla \langle p_0+p_1+p_2 \rangle$$
$$ - {3\bar u^2 \over 12 \eta} \langle h \nabla p_1 \rangle
- {3\bar u \langle h^2 \rangle \over 12 \eta}\nabla p_0 +{1\over 2} \bar u {\bf v}\eqno(3)$$
where we have used that $\langle h \rangle = 0$.
Using that 
$$\nabla \cdot {\bf J} = 0$$ 
we get from (2) to zero order in $h$:
$$\nabla^2 p_0 = 0. $$
The first order contribution gives
$$- {\bar u^3\over 12 \eta} \nabla^2 p_1 - {3\bar u^2 \over 12 \eta} \nabla \cdot (h \nabla p_0) 
+{1\over 2}{\bf v}\cdot \nabla h = 0\eqno(4)$$
We define
$$p_1({\bf q}) = {1\over (2\pi )^2} \int d^2x \ p_1({\bf x}) e^{-i{\bf q}\cdot {\bf x}}$$ 
$$p_1({\bf x}) = \int d^2q \ p_1({\bf q}) e^{i{\bf q}\cdot {\bf x}}$$ 
and similar for $h({\bf x})$.
Substituting these results in (4) gives
$${\bar u^3\over 12 \eta} q^2 p_1 ({\bf q}) - {3\bar u^2 \over 12 \eta} h({\bf q}) (i{\bf q}) \cdot \nabla p_0 
+{1\over 2}{\bf v}\cdot (i{\bf q}) h({\bf q}) = 0\eqno(5)$$
or
$$p_1 ({\bf q}) = 
 {3\over \bar u q^2}  h({\bf q}) (i{\bf q}) \cdot \nabla p_0 
- {6\eta \over \bar u^3 q^2}{\bf v}\cdot (i{\bf q}) h({\bf q})\eqno(6)$$
Next, note that
$$\langle h({\bf q}) h({\bf q'}) \rangle = {1\over (2\pi )^4} \int d^2x d^2x' \ \langle h({\bf x}) h({\bf x'}) \rangle e^{i{\bf q}\cdot {\bf x} +i{\bf q'}\cdot {\bf x'}}$$
$$={1\over (2\pi )^4} \int d^2x d^2x' \ \langle h({\bf x}-{\bf x'}) h({\bf 0}) \rangle e^{i{\bf q}\cdot {\bf x} +i{\bf q'}\cdot {\bf x'}}$$ 
$$={1\over (2\pi )^4} \int d^2x d^2x' \ \langle h({\bf x}-{\bf x'}) h({\bf 0}) \rangle e^{i{\bf q}\cdot ({\bf x}-{\bf x'}) +i({\bf q'}+{\bf q})\cdot {\bf x'}}$$
$${1\over (2\pi )^4} \int d^2x d^2x' \ \langle h({\bf x}) h({\bf 0}) \rangle e^{i{\bf q}\cdot {\bf x} +i({\bf q'}+{\bf q})\cdot {\bf x'}}$$
$$={1\over (2\pi )^2}\int d^2x  \ \langle h({\bf x}) h({\bf 0}) \rangle e^{i{\bf q}\cdot {\bf x}} \delta ({\bf q}+{\bf q'})$$
$$ = C({\bf q})\delta ({\bf q}+{\bf q'})$$
Using this equation and (6) gives
$$\langle h \nabla p_1\rangle = \int d^2q d^2q' \ (i{\bf q'}) \langle h({\bf q}) p_1({\bf q'}) \rangle e^{i({\bf q}+{\bf q'})\cdot {\bf x}}$$ 
$$=\int d^2q \ C({\bf q}) {{\bf q} {\bf q}\over q^2} \cdot \left ({6\eta \over \bar u^3}{\bf v}-  {3\over \bar u} \nabla p_0  \right ) $$ 
Substituting this result in (3) gives
$$\langle {\bf J} \rangle = - {1\over 12 \eta} A(\bar u) \nabla \bar p + {1\over 2} B(\bar u) {\bf v}\eqno(7)$$
where $\bar p = \langle p_0+p_1+p_2\rangle$, and where the $2\times 2$ matrices $A$ and $B$ can be written as $A=\bar u^3 \phi_{\rm p}$ and
$B=\bar u \phi_{\rm s}$ with the flow factor matrices
$$\phi_{\rm p} = 1+{3\over \bar u^2} \left (\langle h^2 \rangle - 3 \int d^2q \ C({\bf q}) {{\bf q} {\bf q}\over q^2} \right )$$
$$= 1+{3\langle h^2 \rangle \over \bar u^2}(1 - 3D), \eqno(8)$$
and 
$$\phi_{\rm s} =1-{3\over \bar u^2} \int d^2q \ C({\bf q}) {{\bf q} {\bf q}\over q^2} = 1-{3\langle h^2\rangle \over \bar u^2} D. \eqno(9)$$
Here we have defined the $2\times 2$ matrix
$$D = {\int d^2q \ C({\bf q}) {\bf q} {\bf q} /q^2 \over \int d^2q \ C({\bf q})}$$
where $q_0$ is the smallest surface roughness wavevector.
For roughness with isotropic statistical properties, $D_{ij} = 1/2$ in which case (8) and (9) becomes
$$\phi_{\rm p}  = \phi_{\rm s} = 1-{3\over 2} {\langle h^2\rangle \over \bar u^2}.\eqno(10)$$
In deriving (7) we have used that to order $h^2$ one can replace terms like $h^2\nabla p_0$ with $h^2\nabla \bar p$. 

In the derivation above we calculated the pressure and shear flow factors to first order in $\langle h^2 \rangle /\bar u^2$. In principle
it is possible to extend the perturbation expansion to calculate higher order terms in $\langle h^2 \rangle /\bar u^2$.
This will result in higher order correlation functions, e.g.,  
$\langle h_1 h_2 h_3 h_4 \rangle$ (where $h_1=h({\bf q}_1)$ and so on),  
but if the surface is randomly rough then these higher order correlation functions
can be decomposed into a sum of products of pair correlation functions, e.g., 
$$\langle h_1 h_2 h_3 h_4 \rangle = \langle h_1 h_2 \rangle \langle  h_3 h_4 \rangle +\langle  h_1 h_3 \rangle \langle  h_2 h_4 \rangle +\langle  h_1 h_4 \rangle \langle  h_2 h_3 \rangle $$
Thus, all terms in the perturbation expansion will only involve the pair correlation function $C({\bf q})$. We empathize that this is the case only
for randomly rough surfaces where the phase of the different plane-wave components in the Fourier decomposition of $h({\bf x})$ are uncorrelated.
However, already the calculation of the second order term in the expansion of the flow factors in $\langle h^2 \rangle /\bar u^2$ becomes
very cumbersome. In Appendix A we present a much simple and more powerful approach, which is in the spirit of the Renormalization Group (RG) procedure.
Thus we eliminate or integrate out the surface roughness components in steps and obtain a set of RG flow equations describing how the
effective fluid equation evolves as more and more of the surface roughness components are eliminated.

Assume that after eliminating  all the surface roughness components with wavevector $|{\bf q}| = q > \zeta q_0$ 
the fluid current takes the form
$${\bf J} = -{1\over 12 \eta} A(u,\zeta) \nabla p + {1\over 2} B(u,\zeta) {\bf v}\eqno(11)$$
where $A$ and $B$ are $2\times 2$ matrices.
In Appendix A we show that $A(u,\zeta)$ and $B(u,\zeta)$ satisfies
$${\partial A \over \partial \zeta} = \left [ {1\over 2} A''(u,\zeta) - A'(u,\zeta) M A'(u,\zeta)\right ]{d\over d\zeta} \langle h^2\rangle_\zeta  \eqno(12)$$
$${\partial B \over \partial \zeta} = \left [ {1\over 2} B''(u,\zeta) - A'(u,\zeta) M B'(u,\zeta)\right ]{d\over d\zeta} \langle h^2\rangle_\zeta \eqno(13)$$
where $A'=\partial A /\partial u$ and so on, and
where the $2\times 2$ matrix $M\sim A^{-1}$ is defined in Appendix A. Here $\langle h^2\rangle_\zeta$ is the mean of the square of the roughness amplitude including only
the roughness components with wavevector $q > \zeta q_0$ which can be written as
$$\langle h^2\rangle_\zeta = \int_{q>\zeta q_0} d^2q \ C({\bf q})\eqno(14)$$
If we assume that $D(\zeta)$ (defined in Appendix A and in Sec. 4) 
is independent of $\zeta$, it is easy to solve these equations using perturbation theory 
to arbitrary order in the surface roughness amplitude $h$. As an example, for random roughness with isotropic statistical properties one 
obtain to second order in $\langle h^2 \rangle /\bar u^2$ (see Appendix A):
$$A = u^3 \left (1-{3\over 2} {\langle h^2\rangle_\zeta\over u^2} -{9\over 8} { \langle h^2\rangle_\zeta^2\over u^4}\right )\eqno(15)$$
$$B = u \left (1 -{3\over 2} {\langle h^2 \rangle_\zeta \over u^2} -{21\over 8} {\langle h^2 \rangle^2_\zeta \over u^4} \right )\eqno(16)$$
The terms to linear order in $\langle h^2 \rangle$ in these expressions agree with the result of Tripp. He compared his expansion
results with the numerical results of Patir and Cheng and found that the expression for $A$ (or $\phi_{\rm p}$) and $B$ (or $\phi_{\rm s}$)
agree rather well with the numerical results for $\langle h^2 \rangle^{1/2}/\bar u < 3$ and $<6$, respectively. For the latter case our 
second order contribution to $B$ improves the agreement 
between numerical results and the expansion result but for $\langle h^2 \rangle^{1/2}/\bar u < 3$
the direct wall-wall interaction becomes so important that the expansion result (which neglect this interaction) cannot be used.

\begin{figure}
\includegraphics[width=0.45\textwidth,angle=0.0]{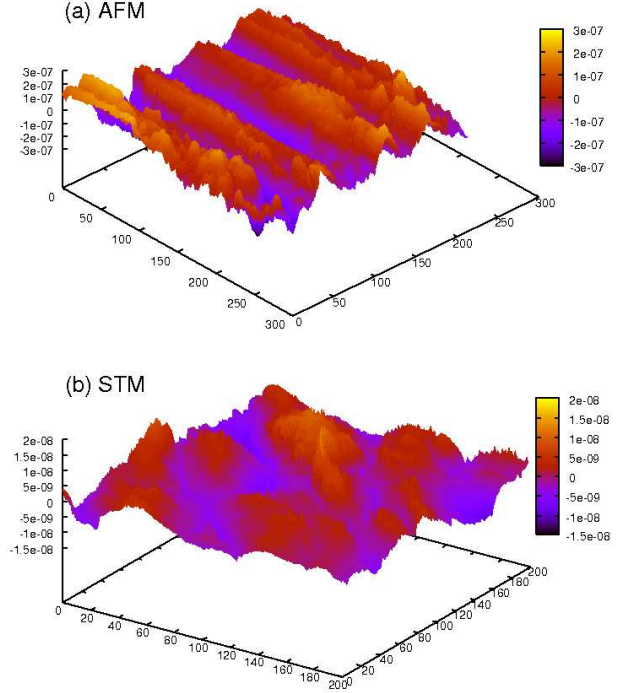}
\caption{\label{picture.grinded.AFM}
Surface topography of a grinded steel surface obtained using
(a) Atomic Force Microscopy (AFM) ($10 {\rm \mu m} \times 10 {\rm \mu m}$) and (b) Scanning Tunneling Microscopy 
(STM) ($0.1 {\rm \mu m} \times 0.1 {\rm \mu m}$).
}
\end{figure}

\begin{figure}
\includegraphics[width=0.45\textwidth,angle=0.0]{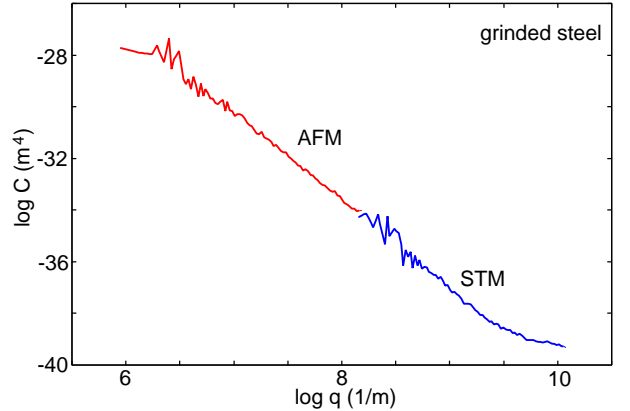}
\caption{\label{Cq.ATM.STM}
The (angular averaged) surface roughness power spectrum $C(q)$ calculated from the AFM and the STM surface topography data 
shown in Fig. \ref{picture.grinded.AFM}. 
}
\end{figure}

\vskip 0.3cm

{\bf 4. Surfaces with anisotropic statistical properties}

As discussed in the introduction, surfaces with anisotropic statistical properties 
are usually characterized by the Peklenik number
$\gamma = \xi_x /\xi_y$, which is the ratio between 
characteristic correlation length $\xi_x$ and $\xi_y$,
defined as the distances along the $x$ and $y$-axis 
where the height-height correlation 
function $\langle h(x,y)h(0,0)\rangle$ has decayed to half of its initial value.
However, for most real surfaces $\gamma (\zeta)$ will depend on the magnification or length-scale under consideration.
Here we propose to obtain $\gamma (\zeta)$ from the surface roughness power spectrum $C(q_x,q_y)$ as follows:

The surface roughness power spectrum $C({\bf q})$ is defined by
$$C({\bf q}) = {1\over (2\pi )^2} \int d^2x \ \langle h({\bf x})h({\bf 0})\rangle e^{i{\bf q}\cdot {\bf x}}$$
We can write
$$C({\bf x}) = \langle h({\bf x})h({\bf 0})\rangle = \int d^2q \ C({\bf q}) e^{-i{\bf q}\cdot {\bf x}}$$
We also define
$$C({\bf x},\zeta) = \int_0^{2\pi} d\phi \ C({\bf q}) e^{-i{\bf q}\cdot {\bf x}}$$
where ${\bf q} = \zeta q_0 ({\rm cos}\phi, {\rm sin}\phi)$.
Now consider the closed contour defined by
$$C({\bf x},\zeta) = C({\bf 0},\zeta)/2$$
We now fit this contour to the quadratic function $f({\bf x}) = a_{ij} x_ix_j + b_i x_i +c$.
The function $a_{ij} x_ix_j = const.$ describes an ellipse which in general has its major axis rotated
by some angle $\psi$ relative to the $x$-axis. We define $\gamma$ as the ratio between the major and minor ellipse 
axis, and obtain both $\gamma (\zeta)$ and the rotation angle $\psi (\zeta)$, both of which depend on the magnification $\zeta$. 

Another way to determine an effective $\gamma (\zeta)$ is as follows: Consider the tensor (see also Appendix A)
$$D(\zeta) = {\int_0^{2\pi} d\phi \ C({\bf q}) {\bf q} {\bf q} /q^2 \over \int_0^{2\pi} d\phi \ C({\bf q})} \eqno(17{\rm a})$$
where ${\bf q} = \zeta q_0 ({\rm cos}\phi, {\rm sin}\phi)$. If $D(\zeta)$ is independent of $\zeta$ then this definition is identical to
$$D = {\int d^2q \ C({\bf q}) {\bf q} {\bf q} /q^2 \over \int d^2q \ C({\bf q})}\eqno(17{\rm b})$$
which appeared already in the perturbation calculation in Sec. 3. 
Note that $D_{11}+D_{22} = {\it Tr} D = 1$ and that
the $D$ is symmetric and can be diagonalized. For example,
suppose $C({\bf q}) = f(\alpha_x q_x^2 + \alpha_y q_y^2)$ and that the ${\bf q}$-integrals in (17b) are over the whole
${\bf q}$-plane. For this case we get after some simplifications
$$D = {1\over 2 \pi} \int_0^{2\pi} d\phi \ {\hat x \hat x {\rm cos}^2 \phi 
+\hat y \hat y \gamma^2 {\rm sin}^2 \phi \over {\rm cos}^2 \phi +\gamma^2 {\rm sin}^2 \phi}\eqno(18)$$
where $\gamma^2 = \alpha_x /\alpha_y$. Performing the integral gives
$D_{11}= 1/(1+\gamma)$ and $D_{22} = \gamma /(1+\gamma)$. Note that in this case $|D|=D_{11}D_{22} = \gamma /(1+\gamma)^2$
where $|D|$ is the determinant of the matrix $D$. This equation has two solutions, $\gamma$ and $1/\gamma$ where
$$\gamma = {1\over 2|D|}\left [1-\left (1-4|D|\right )^{1/2}\right ]-1\eqno(19)$$
Note that this definition of $\gamma$ is independent
of the coordinate system used since the determinant is invariant under rotations (orthogonal transformations).
Note also that for a surface with isotropic statistical properties from (17) $D_{ij} = \delta_{ij} / 2$ so that $|D|=1/4$ and (19)
reduces to $\gamma = 1$ as it should. The angle $\psi$ between the major axis of the ellipse and the $x$-axis of the coordinate system
depends, of course, on the coordinate system, and is given by
$${\rm tan} \psi = c\pm \left (1+c^2\right )^{1/2}\eqno(20)$$
where $c=(D_{22}-D_{11})/(2D_{12})$.

\begin{figure}
\includegraphics[width=0.45\textwidth,angle=0.0]{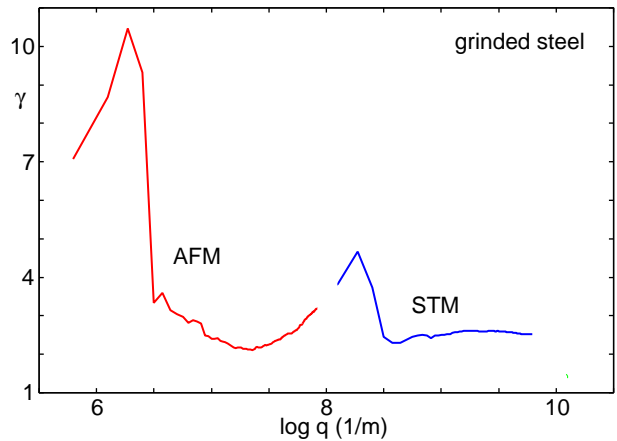}
\caption{\label{gamma.AFM.STM}
The $\gamma$-parameter calculated from the AFM and the STM surface topography data
shown in Fig. \ref{picture.grinded.AFM}. The maximum of $\gamma$ occur for $q \approx 1.8\times 10^6 \ {\rm m}^{-1}$,
corresponding to a wavelength $\lambda = 2 \pi /q \approx 3.5 \ {\rm \mu m}$. This is just the wavelength of
surface topography orthogonal to the major wear tracks in Fig. \ref{picture.grinded.AFM}.  
}
\end{figure}

In Fig. \ref{picture.grinded.AFM} we show the surface topography of a grinded steel surface as obtained using
(a) Atomic Force Microscopy (AFM) ($10 {\rm \mu m} \times 10 {\rm \mu m}$) and (b) Scanning Tunneling Microscopy 
(STM) ($0.1 {\rm \mu m} \times 0.1 {\rm \mu m}$).
In Fig. \ref{Cq.ATM.STM} I show the
(angular averaged) surface roughness power spectrum $C(q)$ calculated from the AFM and the STM surface topography data 
shown in Fig. \ref{picture.grinded.AFM}. 
The power spectrum is well approximated with self affine fractal with the fractal dimension $D_{\rm f} = 2.25$.
However, note that the surface topography is anisotropic. 
In Fig. \ref{gamma.AFM.STM} we show the calculated (using (19)) $\gamma$-parameter for the same surface.
The maximum of $\gamma$ occur for $q \approx 1.8\times 10^6 \ {\rm m}^{-1}$,
corresponding to a wavelength $\lambda = 2 \pi /q \approx 3.5 \ {\rm \mu m}$. This is just the wavelength of
the surface topography orthogonal to the major wear tracks in Fig. \ref{picture.grinded.AFM}.  

\begin{figure}
\includegraphics[width=0.45\textwidth,angle=0.0]{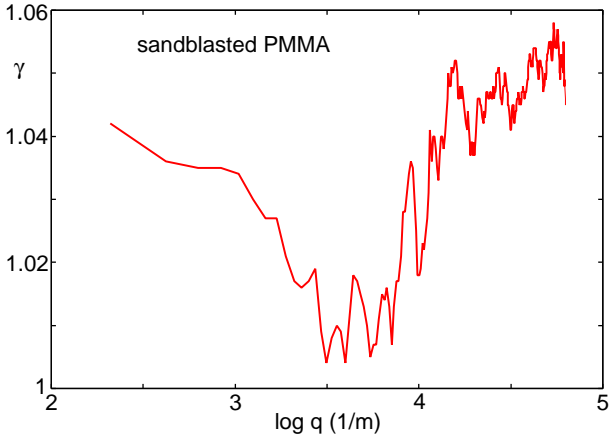}
\caption{\label{sandblastedPMMA.logq.gamma}
The $\gamma$-parameter calculated from optically measured surface topography data for sandblasted
PMMA. The surface topography was measured over a $3 \ {\rm cm} \times 3 \ {\rm cm}$ surface area.
The surface root-mean-square roughness was $32 \ {\rm \mu m}$. 
}
\end{figure}

In Fig. \ref{sandblastedPMMA.logq.gamma} we show the calculated (using (19)) $\gamma$-parameter for a sandblasted
PMMA surface. In this case the statistical properties of the surface are expected to be isotropic, and indeed
$\gamma$ is very close to unity.

For surfaces which have been grinded or polished in one direction, wear scars may occur almost uninterrupted for a very long
distance. In this case it is necessary to measure the surface topography over a very large surface area in order to correctly
obtain the $\gamma(\zeta)$-function. In numerical flow calculations as involved in, e.g., the studies of Patir and Cheng, it would be necessary to
use very large rectangular units which would be practically impossible because of the huge memory and computational time required.

\vskip 0.3cm

\vskip 0.3cm \textbf{5. Contact mechanics: short review and basic equations}

At short (average) interfacial separation there will be a direct asperity interaction between the solids walls,
and in this case the perturbation approach of Sec. 2 will fail. 
Here we will briefly review the contact mechanics model of Persson which we use in this study.

\begin{figure}
\includegraphics[width=0.45\textwidth,angle=0]{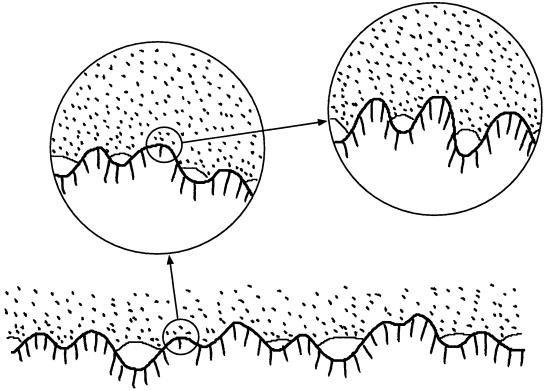}
\caption{\label{1x}
An rubber block (dotted area) in adhesive contact with a hard
rough substrate (dashed area). The substrate has roughness on many different
length scales and the rubber makes partial contact with the substrate on all length scales.
When a contact area
is studied at low magnification it appears as if complete contact occur,
but when the magnification is increased it is observed that in reality only partial
contact occur.
}
\end{figure}

Consider the frictionless
contact between two elastic solids with the Young's elastic modulus $E_0$ and $E_1$ and the Poisson ratios $\nu_0$ and $\nu_1$.
Assume that the solid surfaces have the height profiles $h_0 ({\bf x})$ and $h_1({\bf x})$, respectively. The elastic
contact mechanics for the solids is equivalent to those of a rigid substrate with the height profile $h({\bf x}) = h_0({\bf x})+
h_1({\bf x})$ and a second elastic solid with a flat surface and with the Young's modulus $E$ and
the Poisson ratio $\nu$ chosen so
that\cite{Johnson2}
$${1-\nu^2\over E} = {1-\nu_0^2\over E_0}+{1-\nu_1^2\over E_1}.$$

\begin{figure}
\includegraphics[width=0.35\textwidth,angle=0.0]{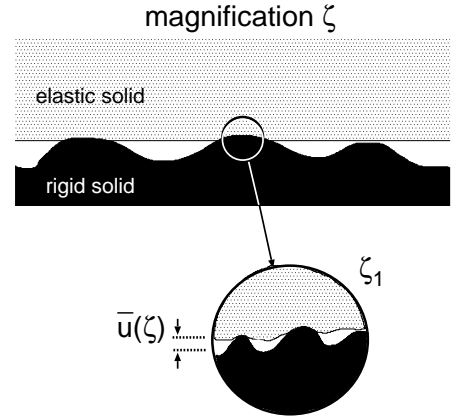}
\caption{\label{asperity.mag}
An asperity contact region observed at the magnification $\zeta$. It appears that
complete contact occur in the asperity contact region, but when the magnification is
increasing to the highest (atomic scale) magnification $\zeta_1$, 
it is observed that the solids are actually separated by the average distance $\bar{u}(\zeta)$.
}
\end{figure}

The contact mechanics formalism developed elsewhere\cite{PSSR,JCPpers,PerssonPRL,YangPersson} is
based on the studying the
interface between two contacting solids at different magnification $\zeta$.
When the system is studied at the magnification $\zeta$ it appears as if the contact area
(projected on the $xy$-plane) equals $A(\zeta)$, but when the magnification
increases it is observed that the contact is incomplete (see Fig. \ref{1x}), and the surfaces in the apparent
contact area $A(\zeta)$ are in fact separated by
the average distance $\bar u(\zeta)$, see Fig. \ref{asperity.mag}.
The (apparent) relative contact area $A(\zeta)/A_0$ at the magnification $\zeta$
is given by\cite{JCPpers,YangPersson}
$${A(\zeta)\over A_0} = {1\over (\pi G )^{1/2}}\int_0^{p_0} d\sigma \ {\rm e}^{-\sigma^2/4G}
= {\rm erf} \left ( p_0 \over 2 G^{1/2} \right )\eqno(21)$$
where
$$G(\zeta) = {\pi \over 4}\left ({E\over 1-\nu^2}\right )^2 \int_{q_0}^{\zeta q_0} dq q^3 C(q)$$
where the surface roughness power spectrum
$$C(q) = {1\over (2\pi)^2} \int d^2x \ \langle h({\bf x})h({\bf 0})\rangle {\rm e}^{-i{\bf q}\cdot {\bf x}}$$
where $\langle ... \rangle$ stands for ensemble average.
The height profile $h({\bf x})$ of the rough surface can be measured routinely
today on all relevant length scales using optical and stylus experiments.

\begin{figure}
\includegraphics[width=0.45\textwidth,angle=0.0]{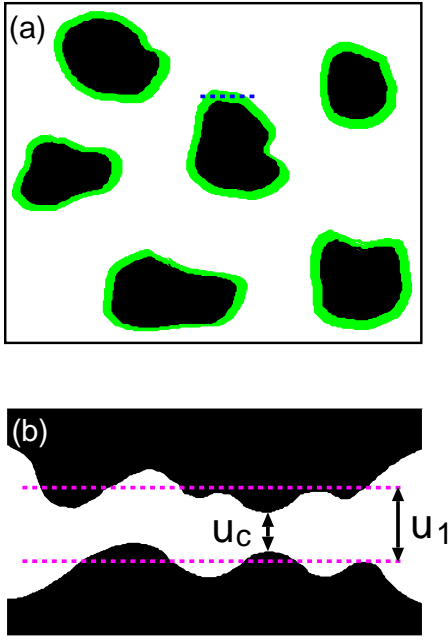}
\caption{\label{pic.Az.Azdz.rough}
(a) The black area is the asperity contact regions at the magnification $\zeta$.
The green area is the additional contact area observed when the magnification is
reduced to $\zeta-\Delta \zeta$ (where $\Delta \zeta$ is small). The average separation
between the solid walls in the green surface 
area is denoted by $u_1(\zeta)$. (b) The separation
between the solid walls along the blue dashed line in (a).
Since the surfaces of the solids are everywhere rough the actual 
separation between the solid walls in the green area
will fluctuate around the average $u_1(\zeta)$. At the most narrow constriction
the surface separation is $u_c$.
}
\end{figure}

The quantity $\bar u(\zeta)$ is the average separation between the surfaces in the apparent contact regions
observed at the magnification $\zeta$, see Fig.~\ref{asperity.mag}. 
It can be calculated from\cite{YangPersson}
$$\bar{u}(\zeta ) = \surd \pi \int_{\zeta q_0}^{q_1} dq \ q^2C(q) 
w(q, \zeta)$$
$$\times \int_{p(\zeta)}^\infty dp' 
 \ {1 \over p'} e^{-[w(q,\zeta) p'/E^*]^2},$$
where $p(\zeta)=p_0A_0/A(\zeta)$
and
$$w(q,\zeta)=\left (\pi \int_{\zeta q_0}^q dq' \ q'^3 C(q') \right )^{-1/2}.$$

We define $u_1(\zeta)$ to be the (average) height separating the surfaces which appear to come into 
contact when the magnification decreases from $\zeta$ to $\zeta-\Delta \zeta$, where $\Delta \zeta$
is a small (infinitesimal) change in the magnification. 
In Fig. \ref{pic.Az.Azdz.rough}(a)
the black area is the asperity contact regions at the magnification $\zeta$.
The green area is the additional contact area observed when the magnification is
reduced to $\zeta-\Delta \zeta$ (where $\Delta \zeta$ is small)\cite{complex}. 
The average separation
between the solid walls in the green surface 
area is given by $u_1(\zeta)$. 
Fig. \ref{pic.Az.Azdz.rough}(b) shows the separation
between the solid walls along the dashed line in Fig. \ref{pic.Az.Azdz.rough}(a).
Since the surfaces of the solids are everywhere rough the actual 
separation between the solid walls in the green area
will fluctuate around the average $u_1(\zeta)$. Thus we expect the smallest surface separation $u_c=\alpha u_1(\zeta_c)$, where
$\alpha < 1$ (but of order unity, see Fig. \ref{pic.Az.Azdz.rough}(b))\cite{WithYang}. 
In Ref. \cite{LP,LPtobe} we have analyzed leak-rate data for rubber seals and always found that $\alpha$
to be in the range $0.5-1$. However, it is clear that $\alpha$ cannot be a fixed constant but must
depend on the average surface separation and on the
surface roughness which occur at length scales shorter than $\lambda = L/\zeta$. In particular,
as $\langle h^2 \rangle_{\zeta}/u_1^2(\zeta) \rightarrow 0$ we expect that $\alpha \rightarrow 1$
(see also Sec. 8). 

$u_1(\zeta)$ is a monotonically decreasing
function of $\zeta$, and can be calculated from the average interfacial separation
$\bar u(\zeta)$ and $A(\zeta)$ using
(see Ref.~\cite{YangPersson})
$$u_1(\zeta)=\bar u(\zeta)+\bar u'(\zeta) A(\zeta)/A'(\zeta).$$
One can show\cite{LP} from the equations above that as the applied squeezing pressure $p_0 \rightarrow 0$,
for the magnifications most relevant for calculating fluid flow (e.g., the leak-rate of seals), $u_1 \rightarrow \bar u$.

We note that when solving for the fluid flow between macroscopic surfaces with roughness one may in a mean-field
type of treatment write the local nominal pressure (i.e., the pressure locally averaged over surface area with linear
dimension of order the wavelength $\lambda_0$ of the longest surface roughness component) as\cite{Scaraggi}
$$p({\bf x},t) = p_{\rm fluid}({\bf x},t)+ p_{\rm solid}({\bf x},t)$$
where $p_{\rm fluid}$ and $p_{\rm solid}$ are locally averaged nominal fluid pressure and solid wall-wall
contact pressure, respectively. The pressure $p_{\rm solid}$ can be related to the interfacial separation
$\bar u({\bf x},t)$ as described in Ref. \cite{PerssonPRL,YangPersson}. In particular, for large enough average surface
separation\cite{PerssonPRL}
$$p_{\rm solid} \approx \beta E^* e^{-\bar u/u_0}$$
where $\beta$ and $u_0$ can be calculated from the surface roughness power spectrum.

\begin{figure}
\includegraphics[width=0.45\textwidth,angle=0.0]{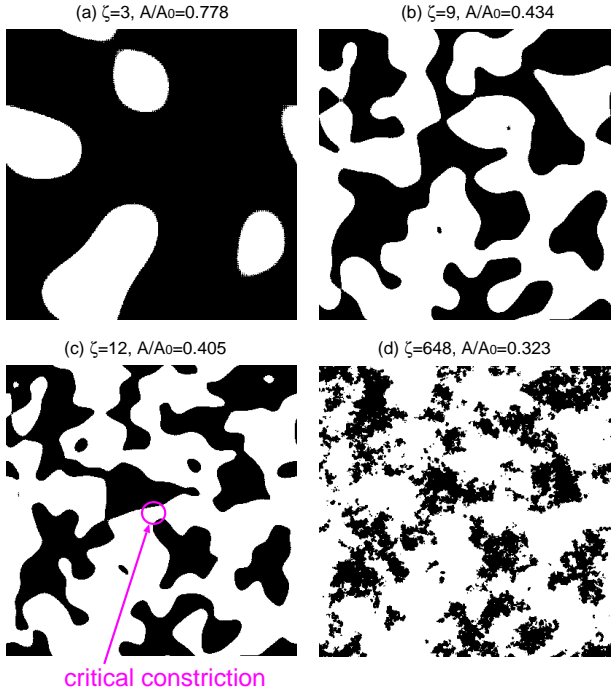}
\caption{\label{percolationpic}
The contact region at different magnifications $\zeta = 3$, 9, 12 and 648, is shown in
(a)-(d) respectively. 
When the magnification increases from 9 to 12 the non-contact region percolate.
At the lowest magnification $\zeta = 1$: $A(1)=A_0$. The figure is the result
of Molecular Dynamics simulations of the contact between elastic solids with randomly rough surfaces,
see Ref. \cite{Yang}.
}
\end{figure}

\vskip 0.3cm
{\bf 6. Critical-junction theory of fluid flow}

The perturbation expansion presented in Sec. 3 assumed no direct contact between the solid walls.
But direct contact between the solid walls occur in most cases of interest, e.g., in static seals.
The simplest approach for this case is based on the leak-rate model developed in Ref. \cite{LP,Creton,P3,Yang,LorenzEPL,Carbone}.
Consider the fluid leakage through a (nominal) contact region, say between a hard solid and rubber,
from a high fluid pressure $P_{\rm a}$ region, to a
low fluid pressure $P_{\rm b}$ region. 
Assume that the nominal contact region between the rubber and the hard countersurface is
rectangular with area $L_x\times L_y$, with $L_y > L_x$. 
We assume that the high pressure fluid region is for $x<0$
and the low pressure region for $x>L_x$. We  ``divide'' the contact region into squares with
the side $L_x=L$ and the area $A_0=L^2$ (this assumes that $N=L_y/L_x$ is an integer, but this 
restriction does not affect the final result). 
Now, let us study the contact between the two solids within one of the squares
as we change the magnification $\zeta$. We define $\zeta= L/\lambda$, where $\lambda$ is the resolution.
We study how the apparent contact area (projected on the $xy$-plane),
$A(\zeta)$, between the two solids depends on the magnification $\zeta$.
At the lowest magnification we cannot observe any surface roughness, and 
the contact between the solids appears to be complete i.e., $A(1)=A_0$. 
As we increase the magnification
we will observe some interfacial roughness, and the (apparent) contact area will decrease.
At high enough magnification, say $\zeta = \zeta_{\rm c}$, a percolating path of 
non-contact area will be observed 
for the first time, see Fig.~\ref{percolationpic}. 
We denote the most narrow constriction along this percolation path as
the {\it critical constriction}. The critical constriction will have the lateral
size $\lambda_{\rm c} = L/\zeta_{\rm c}$ and the surface separation at this point is denoted by 
$u_{\rm c}=\alpha u_1(\zeta_{\rm c})$. 
As we continue to increase the magnification we will find more percolating channels 
between the surfaces, but these will have more narrow constrictions 
than the first channel which appears at $\zeta=\zeta_{\rm c}$, and as a first approximation one may
neglect the contribution to the leak-rate from these channels\cite{Yang}. 

A first rough estimate of the leak-rate is obtained by assuming that all the leakage 
occurs through the critical percolation channel, and that
the whole pressure drop $\Delta P = P_{\rm a}-P_{\rm b}$ (where $P_{\rm a}$ and $P_{\rm b}$ is the 
pressure to the left and right of the
seal) occurs over the critical constriction (of width and length $\lambda_{\rm c} \approx L/\zeta_{\rm c}$
and height $u_{\rm c}$). 
We will refer to this theory as the ``critical-junction'' theory.
If we approximate the critical constriction
as a pore with rectangular cross section (width and length $\lambda_c$ and height $u_c << \lambda_c$),  
and if we assume an incompressible
Newtonian fluid, the volume-flow per unit time through the critical constriction
will be given by (Poiseuille flow) 
$$\dot Q = {u_c^3 \over 12 \eta}  \Delta P,\eqno(22)$$
where $\eta $ is the fluid viscosity. 
In deriving (22) we have assumed laminar flow and that $u_c << \lambda_c$,
which is always satisfied in practice. We have also assumed no-slip boundary condition
on the solid walls. This assumption is not always satisfied at the micro or nano-scale, but is likely to be
a very good approximation in the present case owing to surface roughness which occurs at length-scales 
shorter than the size of the critical constriction. 
Finally, since there are
$N=L_y/L_x$ square areas in the rubber-countersurface (apparent) contact area, we get the total leak-rate
$$\dot Q = {L_y \over L_x} {u_c^3 \over 12 \eta}  \Delta P.\eqno(23)$$ 
Note that a given percolation channel could have several narrow (critical or nearly critical)
constrictions of nearly the same dimension
which would reduce the flow along the channel. But in this case one would also expect more channels from
the high to the low fluid pressure side of the junction, which would tend to increase the leak rate.
These two effects will, at least in the simplest picture where one assumes that the distance between the 
critical junctions along a percolation path (in the $x$-direction) is the same as the distance between the 
percolation channels (in the $y$-direction), compensate 
each other (see Ref. \cite{Yang}). 
The effective medium theory presented below
includes (in an approximate way) all the flow channels.

To complete the theory we must calculate the separation $u_{\rm c}$ 
of the surfaces at the
critical constriction. We first determine the critical magnification $\zeta_{\rm c}$ by assuming that the 
apparent relative contact area at this point is given by percolation theory. 
Thus, the relative contact area $A(\zeta)/A_0 \approx 1-p_{\rm c}$, where $p_{\rm c}$  is the 
so called percolation threshold\cite{Stauffer}. 
For infinite-sized 2D systems, and assuming site percolation,  
$p_{\rm c}\approx 0.70$ for a hexagonal lattice, $0.59$ for a square lattice, and $0.5$ for a triangular lattice\cite{Stauffer}. 
For bond percolation the corresponding numbers are $0.65$, $0.5$, and $0.35$, respectively.
For continuous percolation in 2D 
the Bruggeman effective medium theory predict $p_{\rm c} = 0.5$.
For finite sized systems the percolation will, on the average, occur for (slightly) smaller values
of $p_{\rm c}$, and fluctuations in the percolation threshold will occur between 
different realizations of the same physical system. 
Numerical simulations such as those presented in Ref. \cite{Yang} (see Fig. \ref{percolationpic}) and Ref. \cite{thesis}
typically gives $p_{\rm c}$ slightly larger than $0.5$. In our earlier leak-rate studies we have used
$p_{\rm c} = 0.5$ and $0.6$ to determine the critical magnification $\zeta=\zeta_{\rm c}$. 

We can write the leak-rate in terms of the pressure flow factor. Thus the current
$$J_x = -{\bar u^3 \phi_{\rm p} \over 12 \eta} {d p\over d x} = - {\bar u^3 \phi_{\rm p} \over 12 \eta}{\Delta P \over L_x}$$ 
and the leak-rate
$$\dot Q = J_x L_y = {L_y\over L_x}  {\bar u^3 \phi_{\rm p} \over 12 \eta} \Delta P$$
Comparing this with (23) gives
$$\phi_{\rm p} = \left ({u_c \over \bar u}\right )^3=\left (\alpha {u_1(\zeta_{\rm c}) \over \bar u (1)}\right )^3$$

\begin{figure}
\includegraphics[width=0.4\textwidth,angle=0.0]{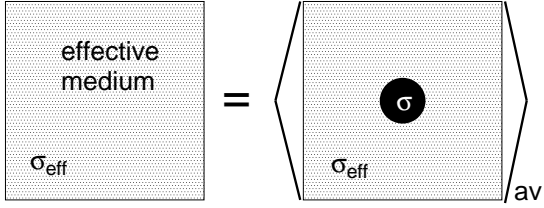}
\caption{\label{effectivemedium}
Effective medium theories take into account random
disorder in a physical system, e.g., random fluctuations in the interfacial
separation $u({\bf x})$. Thus, for a $n$-component system (e.g., where the separation $u$ takes
$n$ different discrete values)
the flow in the effective
medium should be the same as the average fluid flow obtained 
when circular regions of the $n$-components are embedded in the effective medium.
Thus, for example, the pressure $p$ at the origin calculated assuming that the effective medium occur everywhere
must equal the average $\sum c_ip_i$ (where $c_i$ is the concentration of component $i$)
of the pressures $p_i$ (at the origin) calculated with the 
circular inclusion of component $i=1, ..., n$.} 
\end{figure}

\vskip 0.3cm

{\bf 7. Effective medium theory of fluid flow: isotropic roughness}

The critical-junction theory presented above assumes that the leak-rate is determined by the 
resistance towards fluid flow through the critical constriction (or through a network
of critical constrictions, see above). In reality there will be many flow
channels at the interface. Here we will use the 2D Bruggeman effective medium theory\cite{Brugg,Kirk}
to calculate (approximately) the leak-rate resulting from the network of flow channels. Another
approach to extend the critical junction theory is critical path analysis, see Ref. \cite{Bott,Langer}.

We study the
fluid flow through an interface where the separation $u({\bf x})$ between the surfaces varies with
the lateral coordinate ${\bf x} = (x,y)$. 
If $u({\bf x})$ varies slowly with ${\bf x}$ the Navier-Stokes equations of fluid flow reduces to
$${\bf J} = -\sigma \nabla p \eqno(24)$$
where the conductivity $\sigma = u^3({\bf x})/12 \eta$. 

In the effective medium approach one replace the local, spatial varying, conductivity $\sigma ({\bf x})$
with a constant effective conductivity $\sigma_{\rm eff}$. Thus the fluid flow current equation
$${\bf J} = -\sigma_{\rm eff} \nabla p, \eqno(25)$$
as applied to a rectangular region $L_x\times L_y$ with the pressure gradient $dp/dx = (P_{\rm b} - P_{\rm a})/L_x$,
gives
$$\dot Q = L_y J_x = {L_y\over L_x} \sigma_{\rm eff} \Delta P \eqno(26)$$
where $\Delta P = P_{\rm a} - P_{\rm b}$ is the pressure drop.

The effective medium conductivity $\sigma_{\rm eff}$ is obtained as follows.
Let us study the current flow at a circular inclusion (radius $R$) with 
the (constant) conductivity $\sigma$ located
in an infinite conducting sheet with the (constant) conductivity $\sigma_{\rm eff}$. 
We introduce polar coordinates with the origin at the center of
the circular inclusion. The current
$${\bf J} = -\sigma \nabla p \ \ \ \ \ {\rm for} \ \ \ \ \ r<R$$
$${\bf J} = -\sigma_{\rm eff} \nabla p \ \ \ \ \ {\rm for} \ \ \ \ \ r > R$$
We consider a steady state so that
$$\nabla \cdot {\bf J} = 0$$
or
$$\nabla^2 p = 0\eqno(27)$$
If ${\bf J}_0 = - \sigma_{\rm eff} {\bf a}$ is the current far from the inclusion (assumed to be constant) we get
for $r > R$:
$$p = \left [1+f(r)\right ] {\bf a}\cdot {\bf x}\eqno(28)$$ 
Eq. (27) is satisfied if
$$f''(r) + 3 f'(r) r^{-1} = 0$$
A solution to this equation is $f=\alpha r^{-2}$. Substituting this in (28) gives
$$p=\left [1+\alpha r^{-2} \right ] {\bf a}\cdot {\bf x} \eqno(29)$$
For $r < R$ we have the solution
$$p=\beta {\bf a}\cdot {\bf x} \eqno(30)$$
Since $p$ and ${\bf x}\cdot {\bf J}$ must be continuous at $r=R$ we get from (28) and (29):
$$1+\alpha R^{-2} = \beta $$
$$\left (1-\alpha R^{-2} \right ) \sigma_{\rm eff} = \beta \sigma$$
Combining these two equations gives
$$\beta ={2\sigma_{\rm eff} \over \sigma_{\rm eff}+\sigma } \eqno(31)$$
The basic picture behind effective medium theories is presented in Fig. \ref{effectivemedium}.
Thus, for a two component system, 
one assumes that the flow in the effective
medium should be the same as the average fluid flow obtained 
when circular regions of the two components are embedded in the effective medium.
Thus, for example, the pressure $p$ calculated assuming that the effective medium occur everywhere
must equal the average $c_1p_1+c_2p_2$ of the pressures $p_1$ and $p_2$ calculated with the 
circular inclusion of the two components {\bf 1} and {\bf 2}, respectively. For $r < R$
we have for the effective medium $p={\bf a}\cdot {\bf x}$ and using (30) 
the equation $p=c_1p_1+c_2p_2$ gives
$$1= c_1 \beta_1 + c_2 \beta_2\eqno(32)$$
where $c_1$ and $c_2$ are the fractions of the total area occupied by
the components {\bf 1} and {\bf 2}, respectively. Using (31) and (32) gives
$$1= c_1 {2\sigma_{\rm eff} \over \sigma_{\rm eff}+\sigma_1 } + c_2 {2\sigma_{\rm eff} \over \sigma_{\rm eff}+\sigma_2 }$$
which is the standard Bruggeman effective medium for a two component system. Note that if one component is
insulating, say $\sigma_2 = 0$, as $c_1 \rightarrow 0.5$ from above, $\sigma_{\rm eff} \rightarrow 0$, i.e.,
$p_{\rm c} = 1/2$ is the percolation threshold of the two component 2D-Bruggeman effective medium model.

If one instead have a continuous distribution of components (which we number by the continuous index $\xi$) 
with conductivities $\sigma = \sigma (\xi)$, then 
$$1= \int d\xi \ P(\xi) \beta(\xi) \eqno(33)$$
where $P(\xi)$ is the fraction of the total surface area occupied by the 
component denoted by $\xi$. The probability distribution $P(\xi)$ is
normalized so that 
$$\int d\xi \ P(\xi)=1 \eqno(34)$$
Using (31) we get
$$1= \int d\xi \ P(\xi) {2\sigma_{\rm eff} \over \sigma_{\rm eff}+\sigma (\xi) } \eqno(35)$$
It is easy to show from this equation that also for the case of a continuous distribution of
components, the percolation limit occur when the non-conducting component 
(which in our case correspond to the area of real contact where $u=0$ and hence $\sigma = u^3/12\eta = 0$) 
occupies $50 \%$ of the total surface area, i.e., $p_{\rm c} = 1/2$ in this case too.

To summarize, using the 2D Bruggeman effective medium theory we get:
$$\dot Q = {L_y\over L_x} \sigma_{\rm eff} \Delta P, \eqno(36)$$
where $\Delta P= P_{\rm a}-P_{\rm b}$ is the pressure drop and 
where 
$${1\over \sigma_{\rm eff}} = \int d\sigma \ P(\sigma) {2\over \sigma_{\rm eff} + \sigma} $$
$$=  
\int d\zeta \left (-{A'(\zeta)\over A_0 }\right )  {2\over \sigma_{\rm eff} + \sigma (\zeta)}, \eqno(37)$$ 
where
$$\sigma (\zeta ) ={[\alpha u_1(\zeta)]^3 \over 12 \eta}. \eqno(38)$$
Eq. (37) is easy to solve by iteration.

\begin{figure}
\includegraphics[width=0.46\textwidth,angle=0.0]{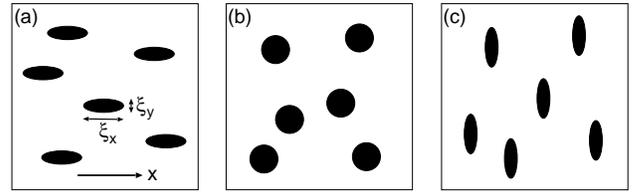}
\caption{\label{gammaLimit}
Contact regions for (a) longitudinal oriented,
(b) isotropic, and (c) transversely oriented rough surfaces. 
The ratio between the ellipse major axis is denoted by $\gamma = \xi_x/\xi_y$ and $\gamma >1$, $=1$ and $<1$ in
(a), (b) and (c), respectively. 
The average fluid flow is
in the $x$-direction.}
\end{figure}

\begin{figure}
\includegraphics[width=0.4\textwidth,angle=0.0]{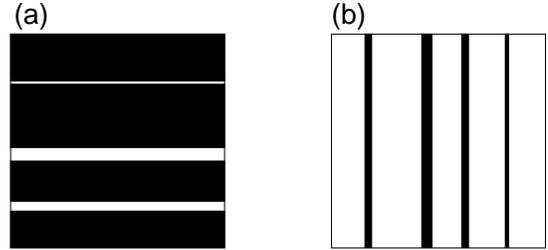}
\caption{\label{twobox}
The black area denote interfacial solid-solid contact 
with the flow conductivity $\sigma_2=0$. The two cases (a) and (b)
correspond to $\gamma = \infty$ and $\gamma = 0$, respectively.
In the first case (a) fluid flow can occur in the strips (open channels) of  
component 1 for arbitrary low concentration of component 1. In this case   
fluid flow will occur at the interface until complete contact occur 
between the solids. In the opposite limit $\gamma \rightarrow 0$ no fluid can 
flow (in the $x$-direction) at the interface unless $c_2$ is zero. 
}
\end{figure}

\vskip 0.3cm

{\bf 8. Effective medium theory of fluid flow: anisotropic roughness}

Here we briefly describe how one may apply the effective medium theory to study fluid flow 
between surfaces with anisotropic (but translational invariant) statistical properties. 
Let $p$ be the locally averages pressure and ${\bf J}$ the fluid flow current also locally averaged.
We have
$$J_i = -\sigma^{\rm eff}_{ij}  {\partial p \over \partial x_j} \eqno(39)$$
Note that
$$\sigma^{\rm eff}_{ij}= {\bar u^3\over 12 \eta} \left (\phi_{\rm p}\right )_{ij}.$$
We can choose a coordinate system such that the flow conductivity tensor is diagonal:

\[
\sigma_{\rm eff} =
\left( {\begin{array}{cc}
 \sigma_\parallel & 0  \\
 0 & \sigma_\perp  \\
 \end{array} } \right)
\]

In this case the $x$ and $y$-coordinate axis are oriented along and perpendicular to the
``groves'' on the surface, respectively. The flow conductivity for any other orientation can be obtained
using the standard transformation of tensors under rotation. Thus if the $x$ axis is oriented an
angle $\phi$ relative to the ``groves'' then

\[
\sigma_{\rm eff} =
\left( {\begin{array}{cc}
\sigma_\parallel {\rm cos}^2 \phi + \sigma_\perp {\rm sin}^2 \phi&  (\sigma_\parallel - \sigma_\perp){\rm cos} \phi{\rm sin}\phi\\
(\sigma_\parallel - \sigma_\perp){\rm cos} \phi{\rm sin}\phi & \sigma_\parallel {\rm sin}^2 \phi +\sigma_\perp  {\rm cos}^2 \phi\\ 
 \end{array} } \right)
\]

We will now calculate the flow conductivities $\sigma_\parallel$ and $\sigma_\perp$ parallel 
and perpendicular to the groves, respectively.
We assume that interfacial separation $u({\bf x})$ varies slowly with ${\bf x} = (x,y)$.
Consider an elliptic inclusion in a fluid. Assume that the fluid flow conductivity equals
$\sigma_{\rm eff}$ outside the inclusion and $\sigma_1$ in the inclusion. 
Assume that the fluid flow far from the
inclusion is oriented at an angle $\phi$ relative to a major axis of the inclusion, 
i.e., far from the inclusion ${\bf J}= - \sigma_{\rm eff} {\bf a}$ and 
$$p={\bf a}\cdot {\bf x} = a(x{\rm cos} \phi +y{\rm sin }\phi)\eqno(40)$$
The fluid flow can be calculated analytically using elliptic coordinates $(\mu, \vartheta)$, see Ref. \cite{MF}. 
In this coordinate system the curves $\mu={\rm const.}$ are ellipses. Consider the
ellipse $\mu = \mu_0$. 
The ratio $\gamma$ between the major and minor axis can be written as $\gamma = {\rm coth}\mu_0$ so that
when $\mu_0 \rightarrow \infty$ the ellipse becomes a circle.  

The fluid pressure inside the elliptic inclusion is given by
$$p= A_{ij}a_ix_j\eqno(41)$$
where the matrix $A_{ij}$ has the components $A_{12}=A_{21}=0$ and
$$A_{11}= {\sigma_{\rm eff} e^{\mu_0} \over \sigma_{\rm eff} {\rm cosh}\mu_0 +\sigma_1 {\rm sinh}\mu_0}\eqno(42)$$
$$A_{22}= {\sigma_{\rm eff} e^{\mu_0}  \over \sigma_1 {\rm cosh}\mu_0 +\sigma_{\rm eff} {\rm sinh}\mu_0}\eqno(43)$$
Note that when $\mu_0 \rightarrow \infty$ the matrix $A_{ij} = \beta \delta_{ij}$, where $\beta$ is given by (A8).
Thus in this case the pressure in the inclusion becomes $p=\beta {\bf a} \cdot {\bf x}$ just as for a circular inclusion
(see Eq. (A7)), which of course is expected because the ellipse becomes a circle when $\mu_0 \rightarrow \infty$. 

The parameter $\mu_0$ is determined by $\gamma = {\rm coth}\mu_0$ or
$$e^{2 \mu_0} = {\gamma+1\over \gamma-1}$$
Using this equation we can also write (42) and (43) as
$$A_{11}= {\sigma_{\rm eff} (\gamma+1) \over \sigma_{\rm eff} \gamma +\sigma_1}\eqno(44)$$
$$A_{22}= {\sigma_{\rm eff} (\gamma+1)  \over \sigma_1 \gamma +\sigma_{\rm eff}}\eqno(45)$$

We now consider the situation where $\phi=0$ so that one of the ellipse axis is oriented 
along the (average) fluid flow direction as in Fig. \ref{gammaLimit}(a).
In this case the pressure in the inclusion
$$p=A_{11} a x$$
while the pressure far away from the inclusion $p=a x$. For a two-component system the effective medium equation (33) now
becomes
$$1={c_1 \sigma_{\rm eff} (\gamma_1+1)\over \sigma_{\rm eff} \gamma_1 +\sigma_1}
+{c_2 \sigma_{\rm eff} (\gamma_2+1) \over \sigma_{\rm eff}\gamma_2 +\sigma_2}\eqno(46)$$
where we have taken into account that the two components may have different ratio $\gamma$. 
Assume that one component, say component 2, has the conductivity $\sigma_2 =0$. In this case it follows from
(46) that $\sigma_{\rm eff} \rightarrow 0$ as $c_2 \rightarrow \gamma_1 /(\gamma_1+1)$. Note in particular that for 
$\gamma_1 \rightarrow \infty$, $c_2 \rightarrow 1$ i.e. in the limit when the major axis of the inclusion 1 goes
to infinite (where conducting strips of the conducting component 1 occur for arbitrary low concentration of component 1)   
fluid flow will occur at the interface until complete contact occur between the solids. In the opposite limit $\gamma_1 \rightarrow 0$,
$c_2 \rightarrow 0$. In this case no fluid can flow (in the $x$-direction) at the interface for any applied pressure. These
two limits correspond to the configurations illustrated in Fig. \ref{twobox}. 

For a continuous distribution of components
$$1=\int d\zeta \ P(\zeta ) {\sigma_{\rm eff} (\gamma(\zeta)+1) \over \sigma_{\rm eff} \gamma(\zeta) +\sigma (\zeta)}\eqno(47)$$
where $\sigma (\zeta) = (\alpha u_1(\zeta))^3/12 \eta$.
This equation is also valid for the orientation of the ellipse as in Fig. \ref{gammaLimit}(c) 
in which case $\gamma < 1$ (in general, $\gamma$ is the ratio between the ellipse axis in the $x$-direction and the $y$-direction).

\begin{figure}
\includegraphics[width=0.45\textwidth,angle=0.0]{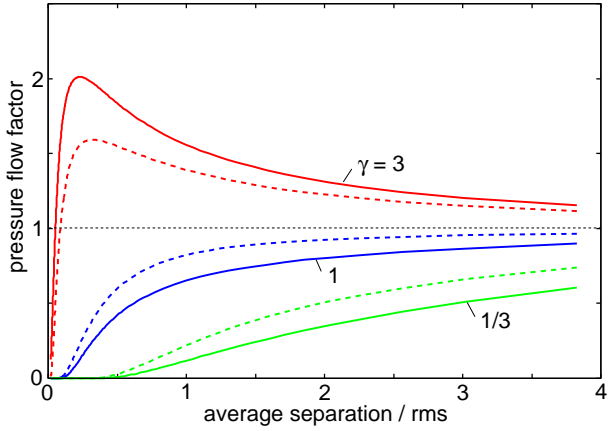}
\caption{\label{PressureFlowFactor.gamm.1.and.3.and.0.333}
The pressure flow factor $\phi_p$ as a function of the average surface separation
$\bar u$ in units of the root-mean-square roughness amplitude. For three different surfaces with 
surface roughness with isotropic statistical properties
($\gamma = 1$), and for surfaces with anisotropic roughness of longitudinal ($\gamma = 3$) and transverse
($\gamma = 1/3$) type. The $\gamma = 1$ case is for sandblasted PMMA (root-mean-square roughness
$22 \ {\rm \mu m}$) in contact with rubber with the elastic modulus $E= 2.3 \ {\rm MPa}$. The other cases
assumes the same angular averaged power spectrum and elastic properties as for the $\gamma = 1$ case.
The solid and dashed lines are discussed in the text.  
}
\end{figure}

\begin{figure}
\includegraphics[width=0.45\textwidth,angle=0.0]{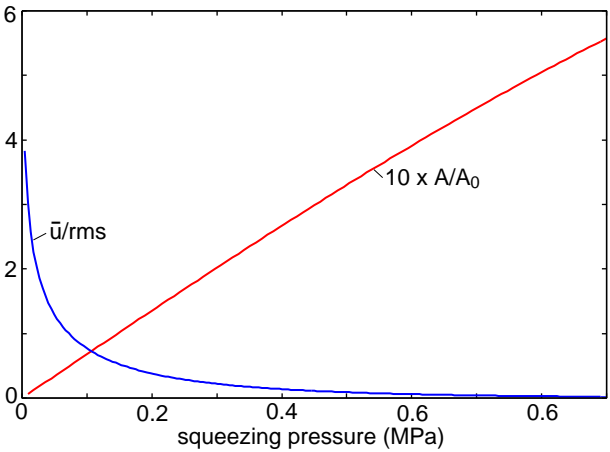}
\caption{\label{1pressure.2Atimes10.and.average.separation}
The variation of the area of real contact $A$ (in units of the nominal contact area $A_0$) and the
average interfacial separation $\bar u$ (in units of the root-mean-square roughness amplitude)
as a function of the (nominal) squeezing pressure for the system 
shown in Fig. \ref{PressureFlowFactor.gamm.1.and.3.and.0.333}: sandblasted PMMA (root-mean-square roughness
$22 \ {\rm \mu m}$) in contact with rubber with the elastic modulus $E= 2.3 \ {\rm MPa}$. 
}
\end{figure}

In Fig. \ref{PressureFlowFactor.gamm.1.and.3.and.0.333} we show the pressure flow factor $\phi_p$ 
as a function of the average surface separation
$\bar u$ in units of the root-mean-square roughness amplitude. 
In the calculation we have for simplicity assumed that $\gamma(\zeta)$ is a constant independent of
the magnification $\zeta$.
Results are shown for
three different surfaces with 
surface roughness with isotropic statistical properties
($\gamma = 1$), and for surfaces with anisotropic roughness of longitudinal ($\gamma = 3$) and transverse
($\gamma = 1/3$) type. The dashed lines is calculated with $\alpha = 1$ while the solid lines is calculated 
with a $\alpha$ which depends on the interfacial separation as follows:   

As pointed out is Sec. 5, the surfaces in the (non-contact) flow channels are everywhere rough, 
and the actual separation between the solid walls in the non-contact region which appears when
the magnification is reduced from $\zeta$ to $\zeta-\Delta \zeta$ 
(green area in Fig. \ref{pic.Az.Azdz.rough}(a)) 
will fluctuate around the average $u_1(\zeta)$. Thus with respect to fluid flow
the separation $u$ between the walls will appear smaller than the average $u_1$ and we use
$u=\alpha u_1(\zeta)$, where
$\alpha < 1$. We note that $\alpha$ is due to the surface 
roughness which occur at length scales {\it shorter}
than $\lambda=L/\zeta$, and it is possible to calculate (or estimate) 
$\alpha$ from the surface roughness power spectrum, as follows.

As shown in Sec. 2 and Appendix A, the fluid flow between two nominal flat
surfaces is affected by the surface roughness on the solid walls even at 
such large (average) surface separation that no direct wall-wall contact occur.
Thus for isotropic roughness at large separation there is a reduction in the fluid flow entering via the
flow factor $\phi_{\rm p} \approx 1-(3/2)(\langle h^2\rangle / u^2)$, where $u$ is the average
surface separation. If we apply this to the present case in the fluid flow problem we replace
the term $u^3_1(\zeta)$ by $u^3_1(\zeta) \phi^*_{\rm p}$ where 
$$\phi^*_{\rm p} = \left ( 1+{3\over 2} {\langle h^2\rangle_ {\zeta^*}\over u_1^2(\zeta^*)} \right)^{-1}$$  
Here we have assumed surface roughness with isotropic statistical properties and
$\langle h^2\rangle_\zeta $ denote the ensemble average of the square of
the roughness amplitude including only the surface roughness with wavevectors larger
than $q=\zeta^* q_0$. In calculating the solid lines in 
Fig. \ref{PressureFlowFactor.gamm.1.and.3.and.0.333} we have chosen $\zeta^* = 3\zeta$. 

Fig. \ref{PressureFlowFactor.gamm.1.and.3.and.0.333} shows, as expected, that when $\gamma$ decreases
the percolation limit, below which no fluid flow can occur, appears at larger and larger average separation. Note also that
for $\gamma =3$ the pressure flow factor first increases with decreasing $\bar u$, but finally it decreases towards zero. Thus,
even for arbitrary large $\gamma$ at high enough squeezing pressures (corresponding to small enough $\bar u$) the
non-contact area  will not percolate in which case no fluid flow can occur at the interface and $\phi_{\rm p} = 0$.

Fig. \ref{1pressure.2Atimes10.and.average.separation}
shows the variation of the area of real contact $A$ (in units of the nominal contact area $A_0$) and the
average interfacial separation $\bar u$ (in units of the root-mean-square roughness amplitude)
as a function of the (nominal) squeezing pressure for the system 
shown in Fig. \ref{PressureFlowFactor.gamm.1.and.3.and.0.333}: sandblasted PMMA (root-mean-square roughness
$22 \ {\rm \mu m}$) in contact with rubber with the elastic modulus $E= 2.3 \ {\rm MPa}$. 
Note that even at the lowest squeezing pressure where $\bar u /rms \approx 4$ the area of real contact is
still non-negligible, about $1\%$ of the nominal contact area.

\vskip 0.3cm
{\bf 9. Summary and conclusion}

I have studied the fluid flow at the interface between elastic solids with randomly rough surfaces.
I have used the contact mechanics model of Persson to take into account the elastic interaction
between the solid walls and the Bruggeman effective medium theory to account for the influence of the disorder on the fluid flow.
I have calculate the flow tensor which determines the pressure flow factor and, e.g., the leak-rate of seals.
I have shown how the perturbation treatment of Tripp
can be extended to arbitrary order in  the ratio between the root-mean-square 
roughness amplitude and the average interfacial surface
separation.  I have introduced a matrix $D(\zeta)$, determined by the surface roughness power spectrum,
which can be used to describe the anisotropy of the surface at any magnification $\zeta$. 
I have present results for the asymmetry factor $\gamma (\zeta)$ (generalized Peklenik number) for
a grinded steel surface and a sandblasted PMMA surface.  

\vskip 0.3cm

{\bf Acknowledgments}

I thank G. Carbone and M. Scaraggi for interesting discussions. I thank A. Wohlers for supplying the AFM and STM
topography data for the grinded steel surface and for discussions.
This work, as part of the European Science Foundation EUROCORES Program FANAS, was supported from funds 
by the DFG and the EC Sixth Framework Program, under contract N ERAS-CT-2003-980409.

\vskip 0.5cm

{\bf Appendix A}

In Sec. 2 we calculated the pressure and shear flow factors to first order in $\langle h^2 \rangle /\bar u^2$. 
Here we will present a simpler and more powerful approach, which is in the spirit of the Renormalization Group (RG) procedure.
Thus we will eliminate or integrate out the surface roughness components in steps and obtain a set of RG flow equations describing how the
effective fluid equation evolves as more and more of the surface roughness components are eliminated.

Assume that after eliminating  all the surface roughness components with wavevector $|{\bf q}| = q > \zeta q_0$ 
the fluid current [given by (1)] takes the form
$${\bf J} = -{1\over 12 \eta} A(u) \nabla p + {1\over 2} B(u) {\bf v}\eqno(A1)$$
where $A$ and $B$ are $2\times 2$ matrices.
We now add to $u$ a small amount of roughness
$$h= \int_{({\zeta-\Delta \zeta}) q_0 < q < \zeta q_0} d^2q \ h({\bf q}) e^{i{\bf q}\cdot {\bf x}}\eqno(A2)$$
Consider now the current
$${\bf J} = -{1\over 12 \eta} A(u+h) \nabla p + {1\over 2} B(u+h) {\bf v}$$
Writing as before
$$p=p_0+p_1+p_2$$ 
we get to second order in $h$
$${\bf J} = - {A(u) \over 12 \eta} \nabla (p_0+p_1+p_2)$$
$$- {A'(u) h \over 12 \eta}\nabla (p_0 +p_1) 
- {A''(u) h^2 \over 24 \eta}\nabla p_0$$
$$ +{1\over 2}(B(u)+B'(u) h) {\bf v}+ {1\over 4}B''(u) h^2 {\bf v}\eqno(A3)$$
The ensemble average of this current gives to second order in $h$
$$\langle {\bf J}\rangle = - {A(u) \over 12 \eta} \nabla \bar p$$
$$- {A'(u) \over 12 \eta}\langle h \nabla p_1 \rangle 
- {A''(u) \langle h^2\rangle \over 24 \eta}\nabla \bar p$$
$$ +{1\over 2}B(u) {\bf v}+ {1\over 4}B''(u) \langle h^2\rangle {\bf v}\eqno(A4)$$
where we have used that $\langle h \rangle=0$.
To zero order in $h$ the continuity equation $\nabla \cdot {\bf J}$ gives
$$A_{ij}(u) \partial_i \partial_j p_0 = 0,$$
and to first order in $h$ we get 
$$-  {A_{ij}'(u) \over 12 \eta} \partial_i h \partial_j p_0 
- {A_{ij}(u) \over 12 \eta} \partial_i\partial_j p_1
 +{1\over 2}B_{ij}'(u) \partial_i h  v_j=0$$
In wavevector space this equation takes the form
$$-  {1 \over 12 \eta} A'_{ij} (u) (iq_i) h({\bf q}) \partial_j p_0 
+ {1  \over 12 \eta} A_{ij}(u) q_i q_j p_1({\bf q})$$
$$ +{1 \over 2}B_{ij}'(u) (iq_i) h({\bf q})  v_j=0$$
or
$$p_1({\bf q}) = (A_{lm}(u) q_l q_m)^{-1} (iq_i) h({\bf q})$$
$$\times    \left ( A_{ij}'(u) \partial_j p_0- 6 \eta B'_{ij} (u)  v_j\right )\eqno(A5)$$
Using this equation and (A2) gives
$$\langle h \partial_i p_1 \rangle = \int d^2q d^2q' \langle h({\bf q}')(iq_i) p_1({\bf q})\rangle $$
$$= \int d^2q \ C(q) (A_{lm}(u) q_l q_m)^{-1}  q_i q_j$$
$$\times \left ( 6 \eta B'_{jk} (u)  v_k-  A_{jk}'(u) \partial_k p_0 \right )\eqno(A6)$$
Let us define the matrix
$$ M_{ij} =  \langle h^2\rangle ^{-1} \int d^2q \ C(q) (A_{lm}(u) q_l q_m)^{-1}  q_i q_j\eqno(A7)$$
so that (A5) becomes
$$\langle h \nabla p_1 \rangle =  6 \eta M B' {\bf v} - M A' \nabla p_0$$
Substituting this in (A4) gives
$$\langle {\bf J}\rangle = - {1\over 12 \eta} \left (A(u) +{1\over 2} \langle h^2\rangle A''(u) - \langle h^2\rangle M A' \right )  \nabla \bar p$$
$$ +{1\over 2}\left ( B(u) + {1\over 2}\langle h^2\rangle B''(u) - \langle h^2\rangle M B'(u) \right ){\bf v}\eqno(A8)$$
Note that this equation has the same general form as the original equation (A1).
If we denote the matrices $A$ and $B$ in the original equation (A1) 
as $A(u,\zeta)$ and $B(u,\zeta)$ to indicate that these where the matrices obtained
after eliminating all wavevector components of $h$ with $q>\zeta q_0$, 
then the new matrices obtained by eliminating the additional roughness 
with wavevectors between $(\zeta-\Delta \zeta)q_0 < q <\zeta q_0$ becomes 
$$A(u, \zeta-\Delta \zeta) = A(u,\zeta) +{1\over 2} \langle h^2\rangle A''(u,\zeta)$$
$$ - \langle h^2\rangle A'(u,\zeta) M A'(u,\zeta)\eqno(A9)$$
$$B(u, \zeta-\Delta \zeta) = B(u,\zeta) + {1\over 2}\langle h^2\rangle B''(u,\zeta)$$
$$ - \langle h^2\rangle A'(u,\zeta) M B'(u,\zeta)\eqno(A10)$$
Since $\Delta \zeta$ is small we can expand the left hand side to linear order in $\Delta \zeta$. Furthermore note that
$${\langle h^2\rangle \over \Delta \zeta} = {1\over \Delta \zeta} \int_{({\zeta-\Delta \zeta}) q_0 < q < \zeta q_0} d^2q \ C({\bf q})$$
$$= {1\over \Delta \zeta} \int_{(\zeta-\Delta \zeta)q_0}^{\zeta q_0} dq q \int_0^{2\pi} d\phi \ C(q{\rm cos}\phi, q{\rm sin}\phi)$$
$$ = \zeta q^2_0  \int_0^{2\pi} d\phi \ C(\zeta q_0{\rm cos}\phi, \zeta q_0{\rm sin}\phi)$$
$$=- {d\over d\zeta} \int_{q>q_0\zeta} d^2q \ C({\bf q}) = - {d\over d\zeta} \langle h^2\rangle_\zeta\eqno(A11)$$
where $\langle h^2\rangle_\zeta$ is the ensemble averaged of the square of the 
roughness amplitude including only roughness with wavevector
$|{\bf q}| > \zeta q_0$.   
Thus from (A9), (A10) and (A11) we get
$${\partial A \over \partial \zeta} = \left [ {1\over 2} A''(u,\zeta) - A'(u,\zeta) M A'(u,\zeta)\right ]{d\over d\zeta} \langle h^2\rangle_\zeta  \eqno(A12)$$
$${\partial B \over \partial \zeta} = \left [ {1\over 2} B''(u,\zeta) - A'(u,\zeta) M B'(u,\zeta)\right ]{d\over d\zeta} \langle h^2\rangle_\zeta \eqno(A13)$$
If we assume that $D(\zeta)$ is independent of $\zeta$, it is easy to solve these equations using perturbation theory 
to arbitrary order in the surface roughness amplitude $h$. 
Since $A\rightarrow u^3$ and $\langle h^2\rangle_\zeta \rightarrow 0$ as $\zeta \rightarrow \zeta_1$ we can write
$$A(u,\zeta) = u^3 +a_1 (u) \langle h^2\rangle_\zeta +a_2(u)  \langle h^2\rangle_\zeta^2 + ...\eqno(A14)$$
To first order in $\langle h^2\rangle_\zeta$ we get from (A12)
$$a_1 = 3u - 9 M  u^4$$
where
$$M = 
{\int_{({\zeta-\Delta \zeta}) q_0 < q < \zeta q_0} d^2q \ C({\bf q}) u^{-3} q^{-2}  {\bf q} {\bf q}\over \int_{({\zeta-\Delta \zeta}) q_0 < q < \zeta q_0} d^2q \ C({\bf q})}$$
or
$$M = u^{-3} {\int_0^{2\pi} d\phi \ C({\bf q}) q^{-2} {\bf q} {\bf q} \over \int_0^{2\pi} d\phi \ C({\bf q})} = u^{-3} D(\zeta)\eqno(A15)$$
where $|{\bf q}|=\zeta q_0$. Thus to first order in $\langle h^2\rangle_\zeta$:
$$A(u,\zeta) = u^3+ \langle h^2\rangle_\zeta u 3(1-3D)$$
$$ = u^3 \left (1+ {\langle h^2\rangle_\zeta \over u^2} 3(1-3D)\right )\eqno(A16)$$
Since $B\rightarrow u$ and $\langle h^2\rangle_\zeta \rightarrow 0$ as $\zeta \rightarrow \zeta_1$ we can write
$$B(u,\zeta) = u +b_1 (u) \langle h^2\rangle_\zeta +b_2(u)  \langle h^2\rangle_\zeta^2 + ...\eqno(A17)$$
Substituting this in (A13) gives
$$b_1 = -3 M  u^2.$$
Thus to first order in $\langle h^2\rangle_\zeta$ we get
$$B(u,\zeta) = u - \langle h^2\rangle_\zeta u^{-1} 3 D = u\left (1-{\langle h^2\rangle_\zeta \over u^2} 3 D\right )\eqno(A18)$$
It is strait forward to calculate the higher order terms (e.g., $a_2$ and $b_2$) in the expansions (A14) and (A17) but here we
will only do so for the case of surface roughness with isotropic statistical properties. In this case $A_{ij} = A(u,\zeta) \delta_{ij}$
and $B_{ij} = B(u,\zeta) \delta_{ij}$. Thus the matrix $M$ in (A7) becomes
$$ M_{ij} =  A^{-1} \langle h^2\rangle ^{-1} \int d^2q \ C({\bf q}) q^{-2}  q_i q_j = {1\over 2} A^{-1} \delta_{ij}$$
and (A12) and (A13) reduces to
$${\partial A \over \partial \zeta} = {1\over 2}\left [ A''(u,\zeta) - {[A'(u,\zeta)]^2 \over A (u,\zeta)}\right ]{d\over d\zeta} \langle h^2\rangle_\zeta  \eqno(A19)$$
$${\partial B \over \partial \zeta} = {1\over 2}\left [ B''(u,\zeta) - {A'(u,\zeta) B'(u,\zeta) \over A (u,\zeta)} \right ]{d\over d\zeta} \langle h^2\rangle_\zeta\eqno(A20)$$
where $A$ and $B$ are now scalar fields. Substituting (A14) in (A19) gives to second order in $\langle h^2\rangle_\zeta$
$$a_1+2a_2 \langle h^2\rangle_\zeta = -{3\over 2} +{1\over 2} \langle h^2\rangle_\zeta \left (a''_1 - 6u^{-1} a'_1 +9 a_1 u^{-2} \right )$$
or
$$a_1 = -{3\over 2} u$$
$$a_2 = {1\over 4} \left (a''_1 - 6u^{-1} a'_1 +9 a_1 u^{-2} \right ) = -{9\over 8} u^{-1}$$
Thus, to second order
$$A = u^3 -{3\over 2} u \langle h^2\rangle_\zeta -{9\over 8} u^{-1} \langle h^2\rangle_\zeta^2$$
$$ = u^3 \left (1-{3\over 2} {\langle h^2\rangle_\zeta\over u^2} -{9\over 8} { \langle h^2\rangle_\zeta^2\over u^4}\right )$$
In a similar way one obtain to second order
$$B = u \left (1 -{3\over 2} {\langle h^2 \rangle_\zeta \over u^2} -{21\over 8} {\langle h^2 \rangle^2_\zeta \over u^4} \right )$$


\begin{thebibliography}{99}
\bibitem{complex} 
Fig. \ref{pic.Az.Azdz.rough}(a) is schematic as in reality
the contact islands at high enough magnification are fractal-like, and decreasing the magnification
result in more complex changes than just adding strips (of constant width) of contact area to 
the periphery of the contact islands. However, this does not change our conclusions.
\end{thebibliography}
\end{document}